# Novel dielectric resonance of composites containing randomly distributed ZrB$_2$ particles with continuous dual-peak microwave absorption

Mengyue Peng, Faxiang Qin[a)]

*Institute for Composites Science Innovation (InCSI), School of Materials Science and Engineering, Zhejiang University, Hangzhou, 310027, China*

**ABSTRACT**

Substantial efforts have been devoted to the elaborate component and microstructure design of absorbents (inclusions) in microwave absorbing (MA) composite materials. However, mesoscopic architectures of composites also play significant roles in prescribing their electromagnetic properties, which are rarely explored in studies of MA materials. Herein, a composite containing randomly distributed ZrB$_2$ particles is fabricated to offer a mesoscopic cluster configuration, which produces a novel dielectric resonance. The resonance disappears and reoccurs when ZrB$_2$ is coated with the insulating and semiconductive ZrO$_2$ layer respectively, suggesting that it is a plasmon resonance excited by the electron transport between ZrB$_2$ particles in clusters rather than any intrinsic resonances of materials constituting the composite. The resonance strength can be regulated by controlling the quantity of the electron transport between particles, which is accomplished by gradually increasing the insulating ZrO$_2$-coated ZrB$_2$ ratio *x* to disturb the electron transport in ternary disordered composites containing ZrB$_2$ and insulating ZrO$_2$-coated ZrB$_2$. When *x* exceeds 0.7, the electron transport is cut off completely and the resonance thus disappears. The resonance induces unusual double quarter-wavelength ( 1/4λ )

[a)] Corresponding author: faxiangqin@zju.edu.cn

interference cancellations or resonance absorption coupled with 1/4λ interference cancellation, giving rise to continuous dual-peak absorption. This work highlights the significance of mesoscopic architectures of composites in MA material design, which can be exploited to prescribe novel electromagnetic properties.

MA materials have attracted increasing attention for their widespread applications in electromagnetic compatibility, wireless communication, and electromagnetic radiation protection.[1-3] According to the frequency dispersion characteristic of electromagnetic constitutive parameters (permittivity and permeability), MA materials can be categorized into relaxation- and resonance-type materials.[4,5] Much attention has been paid to the elaborate component and microstructure design of absorbents (inclusions) in relaxation-type composite materials, obtaining good impedance matching and strong attenuation and hence achieving excellent microwave absorption.[6,7] However, their performance is impeded owing to the single dominant mechanism of interference cancellation.[8] Diversified resonances of resonance-type materials provide abundant resonance absorption mechanisms, expanding the potential of MA performance improvement that deserves more investigation.[9-11] Magnetic resonances are usually restricted to low frequencies as the consequence of Snoek's limit, giving rise to a deteriorative MA performance at higher frequencies.[11,12] As a potential alternative, dielectric resonances can be employed for higher-frequency microwave absorption. However, intrinsic dielectric resonances of molecular and electron only occur in the THz to UV range so that dielectric resonance-type materials receive scant attention in studies of microwave absorbers.[13,14] Therefore, it is necessary but remains a significant challenge to achieve dielectric resonances for microwave absorption.

Here, we demonstrate an effective strategy to construct dielectric resonances at 2-18 GHz by the mesoscopic cluster configuration of composites containing randomly distributed $ZrB_2$ particles. The dielectric resonance is reasoned to be a plasmon resonance induced by electron transport between $ZrB_2$ particles in clusters and its strength can be conveniently modulated by manipulating the quantity of the electron transport. Consequently, uncommon double $1/4\lambda$ interference cancellations or resonance absorption coupled with $1/4\lambda$ interference cancellation induced by the resonance brings about continuous double-peak microwave absorption.

Commercially available $ZrB_2$ particles used in this work are of irregular, stone-like shape with an average size of 2.8 μm and high crystallinity, as shown in Fig. S1. Generally, $ZrB_2$ ceramic exhibits high electric conductivity endowed by its metallic bonding nature.[15,16] Composites containing randomly distributed $ZrB_2$ particles were prepared by homogeneously mixing $ZrB_2$ particles with paraffin at particle concentrations of 60wt.%, 70wt.%, 80wt.%, 85wt.%, and 87wt.%, which were denoted as $ZB_{60}$, $ZB_{70}$, $ZB_{80}$, $ZB_{85}$, and $ZB_{87}$, respectively. One of the important features of disordered composites is the clustering effect of particle fillers, which means some particles are never truly dispersed and isolated from each other, but physically or electrically touch their neighbors, forming particle clusters (agglomerates) resulting from inescapable density fluctuations.[17-22] The electrical transport can take place by tunneling between conductive particles with nanoscale interparticle separation, resulting in electrical touching between nearest particles.[21,23-25] Therefore, these disordered composites are composed of isolated ZrB$_2$ particles and ZrB$_2$ particle clusters embedded in the continuous paraffin. Figures 1(a)-1(c) show scanning electron microscopy (SEM) images of some representative samples (ZB$_{60}$, ZB$_{70}$, and ZB$_{80}$). At the low particle concentration, most ZrB$_2$ particles are surrounded by paraffin, electrically isolated with nearest neighbors, forming isolated particles. As the concentration increases, more clusters emerge from particle agglomeration and they are still electrically isolated from each other.

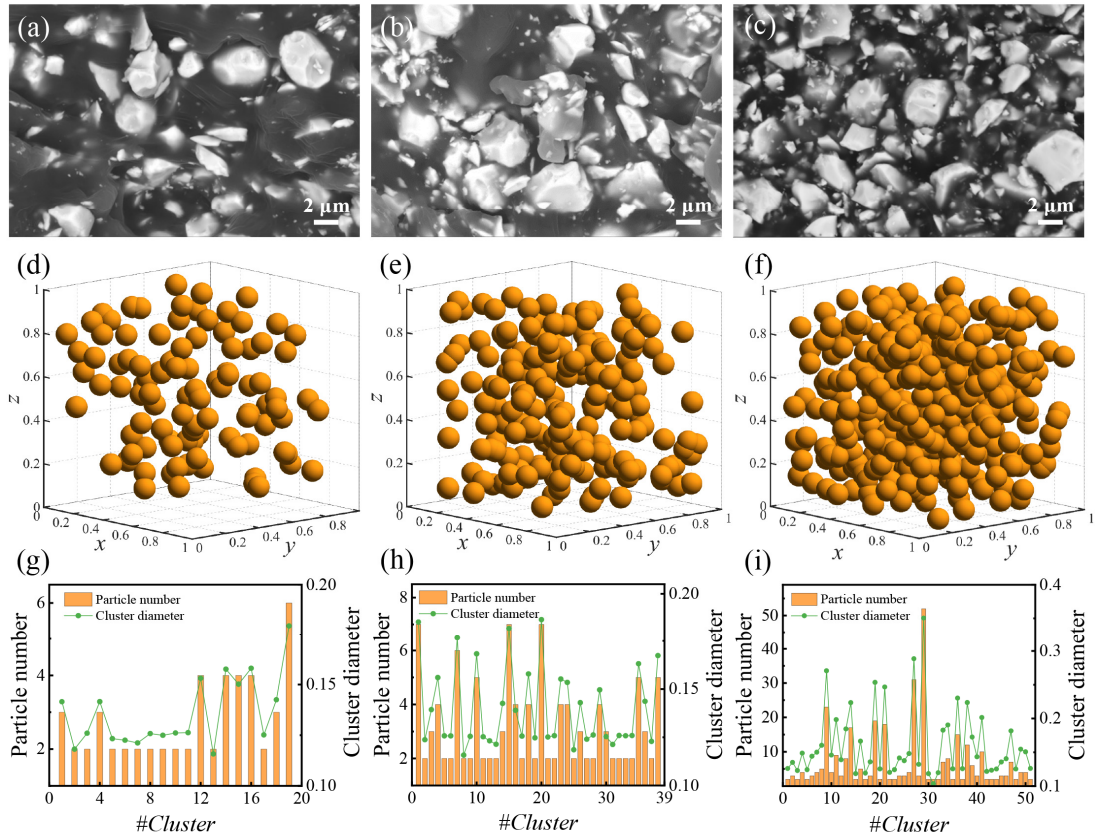

**FIG. 1.** SEM images of (a) ZB$_{60}$, (b) ZB$_{70}$, and (c) ZB$_{80}$. Schematic of disordered composite models with (d) 100, (e) 200, and (f) 400 particles. Particle number and effective diameter of clusters in models with (g) 100, (h) 200, and (i) 400 particles.

Note that isolated clusters become electrically connected and create conductive networks, i.e., percolation, in composites with high ZrB$_2$ contents above 87wt.%.

Further increase of ZrB$_2$ contents to 92wt.% (ZB$_{92}$) and 94wt.% (ZB$_{94}$) would generate negative permittivity resulting from the plasmonic state of free electrons in conductive networks [Fig. S2],[26,27] which will be reported in detail elsewhere.

SEM images of disordered composites cannot disclose the three-dimensional spatial arrangement of clusters. To offer an insightful understanding of cluster behaviors in non-percolating disordered composites, Monte Carlo simulation was implemented to model the agglomeration of particles randomly suspended in the matrix, which generates three-dimensional random distribution of spherical particles with a diameter of 0.1 in a unit cube with the periodic boundary condition and particles can be fully penetrable with each other. The volume fraction of the composite model is equivalent to the total volume occupied by particles. A cluster is defined as the agglomeration of all connected particles. The size of a cluster can be described as the number of particles within the cluster, and the effective diameter of cluster is defined as the diameter of a spherical particle with the same volume as the cluster. Figures 1(d)-1(f) depict disordered composite models with 100, 200, and 400 particles at the volume fraction of 5.07%, 9.94%, and 18.47%, respectively. The particle number and effective diameter of every cluster in these models are demonstrated in Figs. 1(g)-1(i). Obviously, the number and average size of clusters with arbitrary spatial architectures increase as the particle concentration increases. Although the composite model is idealized compared to actual disordered composites, it works for understanding the variation tendency of cluster behaviors.

The relative complex permittivity ($\varepsilon_r = \varepsilon' - j\varepsilon''$) of ZB$_{60}$-ZB$_{87}$ at 2-18 GHz is shown in Figs. 2(a) and 2(b). The $\varepsilon'$ and $\varepsilon''$ values decrease with the increase of frequency at around 2-5 GHz, which is the common dielectric relaxation feature primarily owing to the interfacial polarization between $ZrB_2$ particles and paraffin matrix.[28] More importantly, the $\varepsilon'$ curves exhibit maxima and minima just below and above certain frequencies respectively, whereas sharp peaks on $\varepsilon''$ curves are observed at these frequencies. This is the typical characteristic of dielectric resonance.[29] As $ZrB_2$ content increases, the resonance frequency shifts to lower frequencies, resonance peak amplitude characterizing the resonance strength enhances, and full width at half maximum (FWHM) of the resonance peak expressing the damping condition of resonance increases in a small range [detailed values in Figs. S3(a)-S3(c)].

For a proper measurement and understanding of the resonances of materials, it is necessary to exclude the dimensional resonance excited when the sample thickness is an integer multiple of half wavelength of electromagnetic wave in the material, which is the resonance related to the extrinsic geometrical effect.[30,31] In this condition, sample thickness ($t_D$) satisfies

$$t_D = \frac{nc}{2f\sqrt{|\varepsilon_r \mu_r|}} \ (n = 1,2,3 \ldots), \tag{1}$$

where $c$ is the velocity of electromagnetic wave in vacuum, $f$ is the frequency of electromagnetic wave, and $\mu_r$ is the relative permeability. The actual thickness of $ZB_{60}$-$ZB_{87}$ is around 2 mm, which is smaller than their $t_D$ values of around 5 mm [detailed thickness values in Fig. S3(d)], confirming that the resonance is not dimensional resonance but determined by intrinsic material properties.

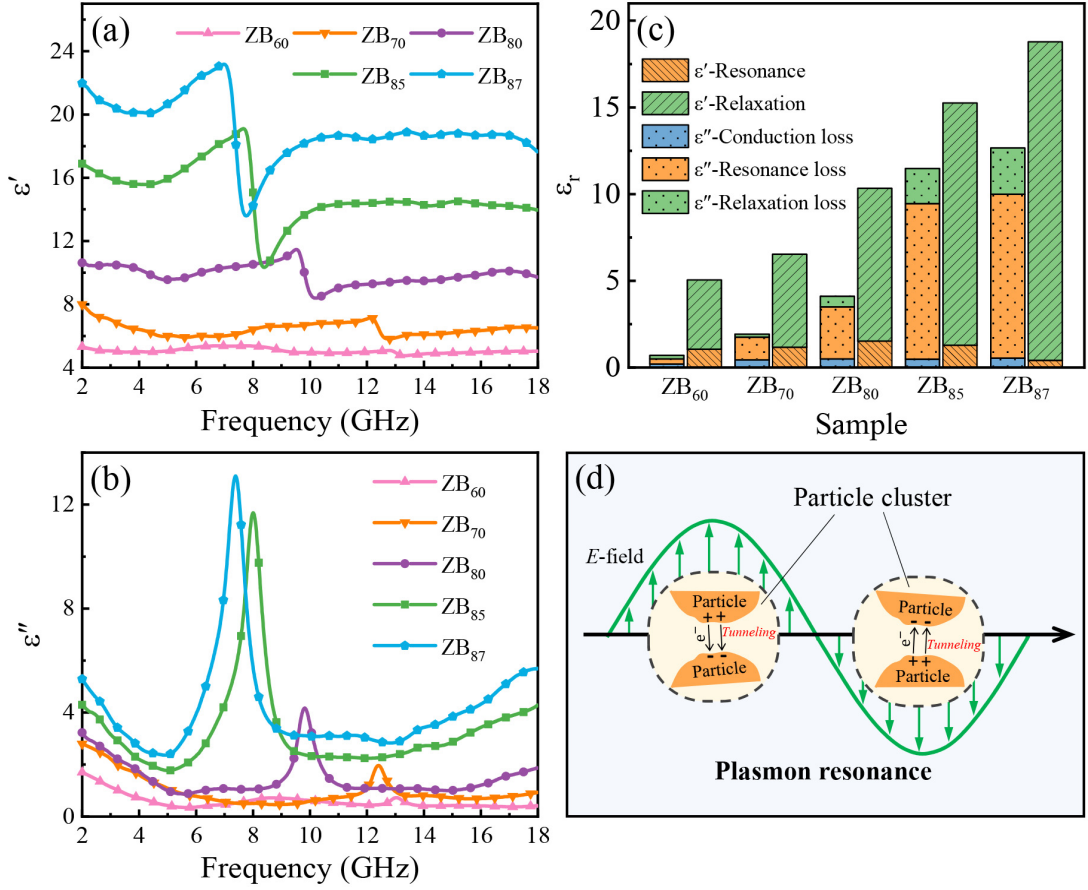

**FIG. 2.** (a) Real and (b) imaginary parts of permittivity of $ZB_{60}$-$ZB_{87}$. (c) Contributions of resonance, relaxation, and conduction loss to permittivity at resonance frequencies. (d) Schematic of plasmon resonance.

As mentioned above, dielectric polarization behaviors of $ZB_{60}$-$ZB_{87}$ include both relaxation and resonance. Moreover, conduction loss might contribute to their dielectric loss. The relaxation and resonance behavior can be well described by Cole-Cole and Lorentz model, respectively.[4,32] Hence, the permittivity of $ZB_{60}$-$ZB_{87}$ can be depicted by a linear combination of the two models and the contribution of conduction loss,[33] namely CLC hybrid model,

$$\varepsilon_r = \varepsilon_\infty + \frac{\varepsilon_s - \varepsilon_\infty}{1+(j\omega\tau)^{(1-\alpha)}} + \frac{\omega_p^2}{\omega_0^2 - \omega^2 + j\Gamma\omega} - j\frac{\sigma}{\omega\varepsilon_0}, \qquad (2)$$

where $\varepsilon_\infty$ and $\varepsilon_s$ are high- and low-frequency permittivity respectively, $\omega(\omega =$

$2\pi f$) is the angular frequency, $\tau$ is the relaxation time, $\alpha$ is an empirical constant with the value between 0 and 1, $\omega_p$ ($\omega_p = 2\pi f_p$) is the plasma angular frequency, $\omega_0$ ($\omega_0 = 2\pi f_0$) is the resonance angular frequency, $\Gamma$ is the damping factor, $\sigma$ is the conductivity, and $\varepsilon_0$ is the permittivity of vacuum. Further, the CLC hybrid model was adopted to fit the measured permittivity of $ZB_{60}$-$ZB_{87}$ assisted by the genetic algorithm (see details in the supplementary material). As shown in Figs. S4(a)-S4(e), numerical fitting results are in good agreement with the experimental data. Relevant parameters ($f_0$, $f_p$, and $\Gamma$) of Lorentz model are shown in Figs. S5(a)-S5(c) and the corresponding complex permittivity is exhibited in Figs. S5(d) and S5(e). At frequencies away from resonance frequencies, the dielectric polarization of $ZB_{60}$-$ZB_{87}$ is relaxation and the dielectric loss is attributed to relaxation loss and conduction loss. However, the dielectric polarization includes both relaxation and resonance and the dielectric loss is composed of resonance loss, relaxation loss, and conduction loss near resonance frequencies. Figure 2(c) shows individual contributions of relaxation, resonance and conduction loss at resonance frequencies of $ZB_{60}$-$ZB_{87}$, indicating that the dielectric polarization and loss at resonance frequencies are dominated by relaxation and resonance, respectively.

The mechanism of the resonance is further elucidated. From the perspective of components, these composites are merely composed of conductive, microwave-reflecting $ZrB_2$ particles and insulating, microwave-transmitting paraffin with permittivity of 2.25. The negative permittivity of $ZB_{92}$ and $ZB_{94}$ indicates that conductive $ZrB_2$ cannot have intrinsic microwave dielectric resonances. Therefore, the resonance cannot arise from intrinsic resonances of materials constituting the composite. From the perspective of mesoscopic configurations of composites, $ZrB_2$ microparticles randomly distributed in the paraffin are in two forms of isolated particles and particle clusters. Generally, isolated conductive particles in composites contribute to the microwave dielectric relaxation owing to the interfacial polarization between them and the matrix. In $ZrB_2$ particle clusters, physically connected particles could be considered as a larger particle, whereas electron transport can be supported by tunneling between separated particles. Hence, it is safely inferred that the resonance is closely related to the electron transport between $ZrB_2$ particles in clusters, which is confirmed experimentally as follows.

$ZrB_2$ particles coated with $ZrO_2$ were prepared via slight oxidation of $ZrB_2$ at 650℃ for 0.5 and 1 h, which were denoted as ZZ-1 and ZZ-2, respectively. ZZ-1 and ZZ-2 were annealed at 800℃ for 1 h in the argon atmosphere, which were named ZZ-1 anneal and ZZ-2 anneal, respectively. ZZ-1 and ZZ-2 are covered by $ZrO_2$ nanoflakes [Fig. S6], whereas the surface of ZZ-1 anneal and ZZ-2 anneal are stacked $ZrO_2$ nanoparticles [Figs. 3(a) and 3(b)]. Additionally, SEM images of well-ground ZZ-1 anneal and ZZ-2 anneal particles demonstrate pronounced core-shell structures and their $ZrO_2$ shell thickness is around 100 and 300 nm respectively [inset of Figs. 3(a) and 3(b)]. Weak steamed bread peaks at around $2\theta = 30°$ in X-ray diffraction patterns (XRD) of ZZ-1 and ZZ-2 indicate majority of their $ZrO_2$ shells are amorphous, whereas new diffraction peaks of ZZ-1 anneal and ZZ-2 anneal show their $ZrO_2$ shells are crystalline [Fig. 3(d)]. Disordered composites containing ZZ-1, ZZ-2, ZZ-1 anneal, and ZZ-2 anneal dispersed in paraffin at the particle content of 85wt.% were denoted as (ZZ-1)$_{85}$, (ZZ-2)$_{85}$, (ZZ-1 anneal)$_{85}$, and (ZZ-2 anneal)$_{85}$, respectively. Permittivity spectra of these composites are shown in Figs. 3(e) and 3(f). Interestingly, (ZZ-1)$_{85}$ and (ZZ-2)$_{85}$ only exhibit relaxation behavior with the disappearance of resonance, whereas the resonance with dramatically reduced strength reoccurs in (ZZ-1 anneal)$_{85}$ and (ZZ-2 anneal)$_{85}$.

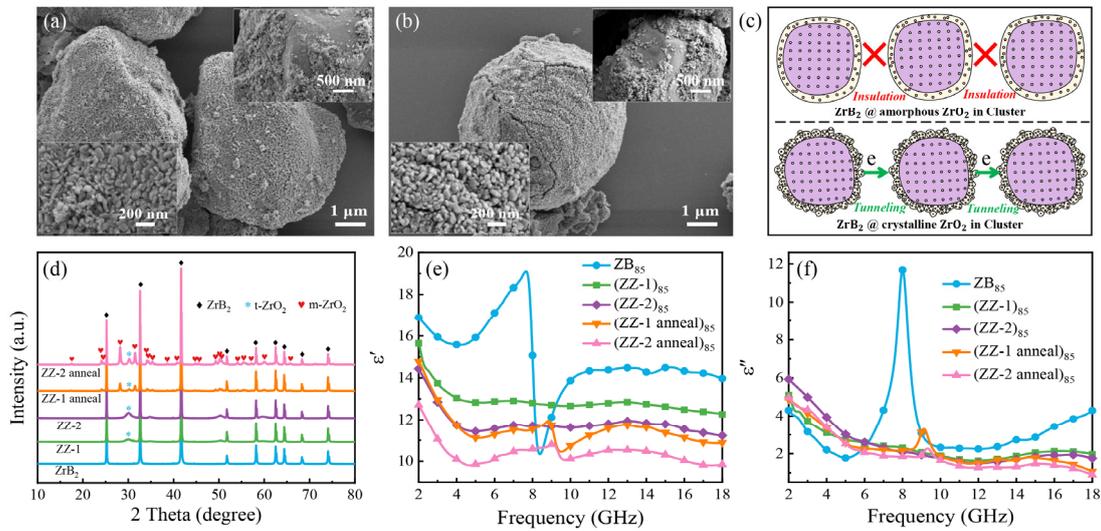

**FIG. 3.** SEM images of (a) ZZ-1 anneal and (b) ZZ-2 anneal. (c) Schematic of electron transport between ZrO$_2$-coated ZrB$_2$ particles. (d) XRD patterns of particles. (e) Real and (f) imaginary parts of permittivity of composites.

The critical difference between (ZZ-1)$_{85}$/(ZZ-2)$_{85}$ and (ZZ-1 anneal)$_{85}$/(ZZ-2 anneal)$_{85}$ is the crystallization of ZrO$_2$ shell. The amorphous ZrO$_2$ suffering from poor carrier concentration and electron mobility exhibits substantial deterioration in electrical conductivity compared to the crystalline ZrO$_2$.[34-36] Additionally, a slight quantity of oxygen vacancies may occur in the crystalline ZrO$_2$ shell of (ZZ-1 anneal)$_{85}$/(ZZ-2 anneal)$_{85}$ during the annealing, enhancing the conductivity of crystalline ZrO$_2$.[37,38] Hence, the amorphous and crystalline ZrO$_2$ shell could be considered as an insulator and semiconductor, respectively. Consequently, the insulating $ZrO_2$ shell with several-hundred-nanometer thickness results in the absence of resonance in $(ZZ-1)_{85}/(ZZ-2)_{85}$. Considering the insulating $ZrO_2$ shell as a part of the insulating paraffin matrix, $(ZZ-1)_{85}/(ZZ-2)_{85}$ containing only isolated $ZrB_2$ particles merely exhibits relaxation behavior, indicating that the resonance is determined by $ZrB_2$ particle clusters rather than isolated $ZrB_2$ particles. Moreover, the electron transport between insulating $ZrO_2$-coated $ZrB_2$ particles in clusters is cut off completely, which should be responsible for the absence of resonance. However, the electron transport can recover partly when $ZrB_2$ is coated with the semiconductive $ZrO_2$ so that the resonance with reduced strength reoccurs in $(ZZ-1\ anneal)_{85}/(ZZ-2\ anneal)_{85}$. Figure 3(c) shows the schematic of electron transport between $ZrO_2$-coated $ZrB_2$ particles in clusters.

Therefore, the resonance is determined by the electron transport between particles in clusters. The more electron transport is supported, the stronger the resonance will be. The localized electrons between particles will collectively oscillate under the action of external electrical field and internal elastic restoring force tending to return them into non-polarized states. When the frequency of electric field is equal to the collective oscillation frequency of electrons, plasmon resonance is excited, as shown in Fig. 2(d). Different from the localized surface plasmon resonance (LSPR) of nanoscale metals with high free electron densities in the visible region,[13] the plasmon resonance in the microwave range is ascribed to low-density localized electrons between $ZrB_2$ particles rather than high-density free electrons inside $ZrB_2$ particles. $ZrB_2$ particles behave as "electron sources" that provide electrons for tunneling between particles. The plasmon resonance might be excited in disordered composites with other conductive particles with appropriate carrier concentrations, which needs exploration in the future.

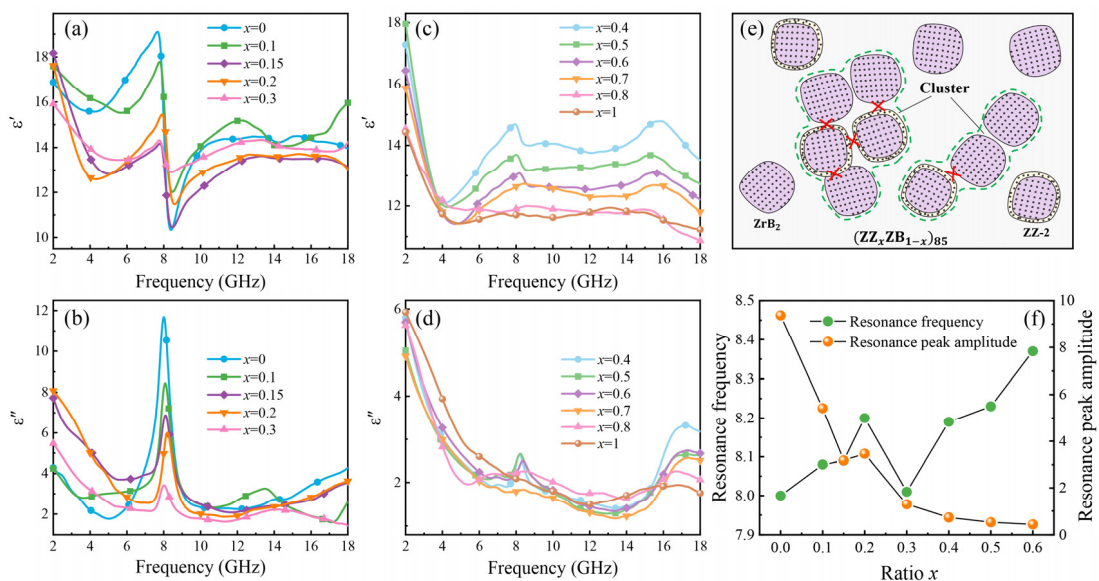

**FIG. 4.** (a, c) Real and (b, d) imaginary parts of permittivity of $(ZZ_xZB_{1-x})_{85}$ ($x$=0-1). (e) Schematic of electron transport interruption in $(ZZ_xZB_{1-x})_{85}$. (f) Resonance frequency and resonance peak amplitude of $(ZZ_xZB_{1-x})_{85}$ ($x$=0-0.6).

The resonance strength is further modulated by controlling the quantity of electron transport between particles in ternary disordered composites containing $ZrB_2$ and insulating $ZrO_2$-coated $ZrB_2$ particles (ZZ-2) dispersed in paraffin at the total particle content of 85wt.%. The composites were denoted as $(ZZ_xZB_{1-x})_{85}$ ($x$=0-1) where $x$ represents the mass ratio of ZZ-2 to total particles. ZZ-2 acts like a "switch" that controls the quantity of electron transport between particles since any electron transport between particles with ZZ-2 is interrupted, as shown in Fig. 4(e). Therefore, electron transport is gradually switched off as ZZ-2 ratio $x$ increases. Figures 4(a)-4(d) show permittivity spectra of $(ZZ_xZB_{1-x})_{85}$ ($x$=0-1). Obviously, the resonance strength gradually decreases and resonance frequency tends to increase in a small frequency range as the ratio $x$ increases until the resonance disappears when $x$ exceeds 0.7 [Fig. 4(f)], indicating that the electron transport in composites is cut off completely when $x \geq 0.7$.

Finally, microwave absorption performance based on the studied resonance is investigated. Reflection loss (RL) is employed to evaluate the microwave absorption performance of composites, which can be expressed by[39]

$$\text{RL} = 20\log\left|\frac{Z_{\text{in}}-Z_0}{Z_{\text{in}}+Z_0}\right|, \quad (3)$$

$$Z_{\text{in}} = Z_0\sqrt{\frac{\mu_r}{\varepsilon_r}}\tanh\left(j\frac{2\pi fd}{c}\sqrt{\mu_r\varepsilon_r}\right), \quad (4)$$

where $Z_{\text{in}}$ represents the input impedance and $Z_0$ is the characteristic impedance of free space. Figures 5(a)-5(c) show RL curves of representative $(ZZ_xZB_{1-x})_{85}$ ($x$=0, 0.15, 0.3) with different resonance strength. $ZB_{85}$ ($x$=0) exhibits dual absorption peaks with a gap of insufficient absorption and the minimum RL value of -48.8 dB at 2.4 mm, whereas a single broad bandwidth is generated by continuous dual-peak absorption without gap and the effective absorption bandwidth (RL≤-10 dB) is 2.3 GHz at 2.7 mm when $x$=0.15. A narrow, weak absorption peak is observed when $x$=0.3. Although continuous dual- or multi-peak absorption is common in metamaterials constructed by multiple resonance elements, such phenomenon is rare in ordinary MA materials.[40,41]

The mechanism of absorption peaks is explored to reveal the generation of dual-peak absorption. Interference cancellation is the most common mechanism of absorption peaks for MA materials with thickness $d$, which can be described by[8,42]

$$d = \frac{nc}{4f\sqrt{|\varepsilon_r\mu_r|}} \quad (n = 1,3,5,\dots). \quad (5)$$

Matching thickness curves of $1/4\lambda$ interference cancellation ($d$-$1/4\lambda$) for $(ZZ_xZB_{1-}$

$_x$)$_{85}$ ($x$=0, 0.15, 0.3) are demonstrated in Figs. 5(d)-5(f) where Y and X-axis of colored dot symbols represent corresponding sample thickness ($d$-sample) and frequencies of absorption peaks of RL curves with the same color, respectively. Some dots are located near the $d$-1/4λ curve and shift to lower frequencies with the increase of sample thickness, confirming that corresponding absorption peaks are mainly attributed to 1/4λ interference cancellation, whereas other dots are away from the $d$-1/4λ curve and near the resonance frequency, indicating that corresponding absorption peaks result from the dielectric resonance. Specifically for ZB$_{85}$, both absorption peaks are ascribed to 1/4λ interference cancellation at 2.0-2.6 mm, whereas lower- and higher-frequency absorption peaks originate from 1/4λ interference cancellation and resonance absorption respectively at 2.7-4.0 mm. Hence, double 1/4λ interference cancellations or combination of the dielectric resonance absorption and 1/4λ interference cancellation is responsible for the dual-peak absorption of (ZZ$_x$ZB$_{1-x}$)$_{85}$ ($x$=0, 0.15).

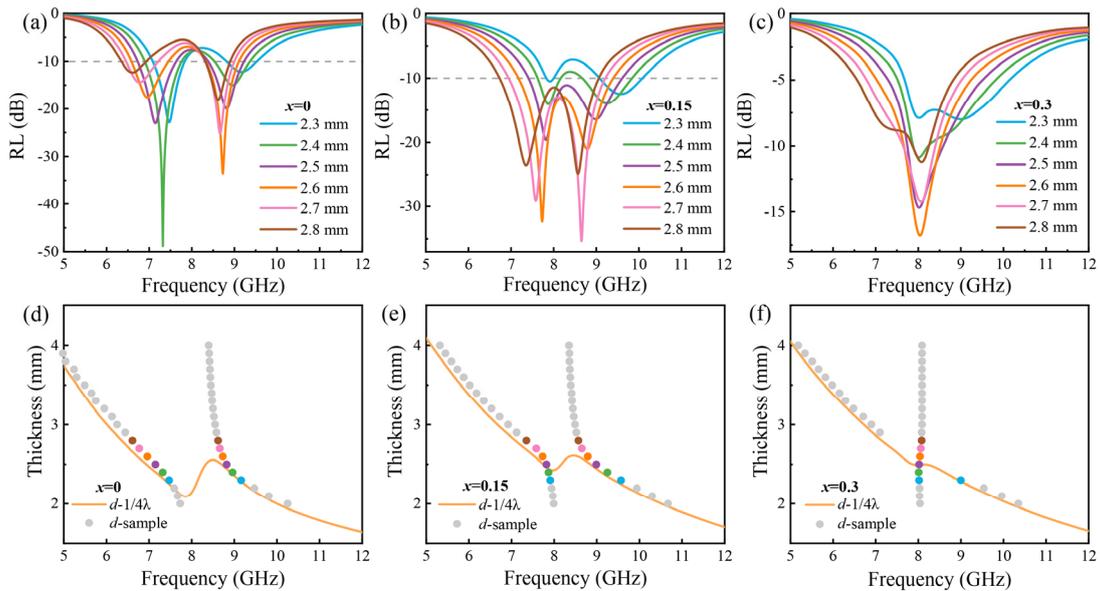

**FIG. 5.** (a, b, c) RL curves, and (d, e, f) matching thickness of 1/4λ interference cancellation and sample thickness of (ZZ$_x$ZB$_{1-x}$)$_{85}$ ($x$=0, 0.15, 0.3).

Note that only one $1/4\lambda$ interference cancellation is usually excited in relaxation-type MA materials with the given thickness for their gentle variation of permittivity and permeability over a broad frequency range,[5,43] whereas the dramatic variation of permittivity near the resonance frequency can enable double $1/4\lambda$ interference cancellations simultaneously for $(ZZ_xZB_{1-x})_{85}$ ($x$=0, 0.15). Moreover, $(ZZ_{0.15}ZB_{0.85})_{85}$ with appropriate resonance strength possesses relatively better impedance matching and adequate attenuation, generating a single broad bandwidth merged by two continuous absorption peaks.

In summary, dielectric behaviors of resonance, relaxation, and negative permittivity at 2-18 GHz were constructed via mesoscopic architectures of composites containing randomly distributed $ZrB_2$ particles. The rare resonance and negative permittivity are determined by mesoscopic cluster and network configurations, respectively. The resonance of concern disappears and reoccurs when $ZrB_2$ is coated with the insulating and semiconductive $ZrO_2$ layer respectively, elucidating that isolated $ZrB_2$ particles contribute to the relaxation, whereas the electron transport between $ZrB_2$ particles in clusters excites plasmon resonance that is responsible for the resonance. The resonance strength is regulated by controlling the quantity of electron transport between particles, which is achieved by gradually increasing the insulating $ZrO_2$-coated $ZrB_2$ ratio $x$ to interrupt the electron transport in ternary disordered composites containing $ZrB_2$ and insulating $ZrO_2$-coated $ZrB_2$ until the electron transport is cut off completely and the resonance thus disappears when $x$ exceeds 0.7. Consequently, unusual double $1/4\lambda$ interference cancellations or resonance absorption coupled with $1/4\lambda$ interference cancellation induced by the resonance generates continuous dual-peak absorption. The effective modulation of dielectric dispersion and further achievement of special microwave absorption from the perspective of composites science that centers on the mesoscopic structure-property relationship suggest the mesostructure design is crucial to bring about novel electromagnetic properties, which are worth more research efforts.

See the supplementary material for experimental details; SEM images; XRD patterns; negative permittivity spectra; characteristic parameters of the dielectric resonance; fitted permittivity spectra; and fitting parameters.

This work is supported by National Key Research and Development Program of China No. 2021YFE0100500, 2021YFB3501504 and Zhejiang Provincial Key Research and Development Program (2021C01004, 2024C01157).

# AUTHOR DECLARATIONS

## Conflict of Interest

The authors have no conflicts to disclose.

## Author Contributions

**Mengyue Peng**: Conceptualization (equal); Investigation (equal); Methodology (equal); Writing – original draft (equal). **Faxiang Qin**: Conceptualization (equal); Supervision (lead); Funding acquisition (lead); Writing – review & editing (equal).

# DATA AVAILABILITY

The data that support the findings of the study are available from the corresponding author upon reasonable request.